\documentstyle[12pt,epsf]{article}
\begin{document}


\begin{center}{\bf Vacuum energy in a spherically symmetric background field}\\[24pt]

by {Michael Bordag\footnote{E-mail address: 
bordag@qft.physik.uni-leipzig.d400.de} and 
Klaus Kirsten\footnote{E-mail address:
kirsten@tph100.physik.uni-leipzig.de}  \\
Universit{\"a}t Leipzig, Institut f{\"u}r Theoretische Physik,\\ 
Augustusplatz 10, 04109 Leipzig, Germany}

\end{center}
\begin{abstract}
The vacuum energy of a scalar field in a spherically symmetric background field
is considered. It is expressed through the Jost function of the corresponding
scattering problem. The renormalization is discussed in detail and performed
using the uniform asymptotic expansion of the Jost function. The method is
demonstrated in an simple explicit example.
\end{abstract}

Pacs number(s): 11.10-z; 02.30.-f; 11.10Gh

hep-th/9608070

\section{Introduction}
The evaluation of quantum corrections to classical solutions play an
important role in several areas of modern theoretical physics. The
classical solutions involved may be monopoles 
\cite{thooft74,polyakov74}, sphalerons
\cite{klinkhamermanton84} and electroweak skyrmions
\cite{gipsontze81,gipson84,ambjornrubakov85,eilamklabucarstern86}
\cite{friedberglee77,friedberglee77a,skyrme61,skyrme72,adkinsnappiwitten83}.
In general the classical fields are inhomogeneous configurations. Thus,
as a rule, the effective potential approximation to the effective action,
where quantum fluctuations are integrated out about a constant classical
field, is not expected to be adequate. The derivative expansion 
\cite{chan85} improves on this by accounting for spatially
varying background fields, being a perturbative approximation it has
however its own limitations. Having in mind that even the classical
solutions are often known only numerically, it is clear that it is
desirable to have a numerical procedure to determine the quantum
corrections. Some efforts in this direction have already been 
undertaken \cite{baacke90,baackekiselev93,lee94,brahmlee94}.

The aim of the present article is to develop further a regular analytic
approach which reduces the evalution of quantum corrections to the 
corresponding quantum mechanical scattering problem. This approach has
been developed for theories in $(1+1)$ dimensions 
\cite{bochkarev92,bordag95}, however, the difficult problem of the
summation over the angular momentum necessary in $(3+1)$ dimensions  
with spherical symmetry has not been  
addressed there. It is thus the aim of the present article to develop
further the regular analytic approach and apply it to theories in 
the physically most interesting $(3+1)$-dimensional spacetime.
Connected with the angular momentum sum, some aspects of the renormalization
are also more complicated here and are discussed in detail.
\section{The model and its renormalization}
Let us first describe our concrete model and its renormalization introducing
also various notations used in the following. We will 
consider the Lagrangian 
\begin{eqnarray}
L=\frac 1 2 \Phi (\Box -M^2 -\lambda \Phi ^2 )\Phi 
+\frac 1 2 \varphi (\Box -m^2 -\lambda ' \Phi^2 )\varphi .\label{22.1}
\end{eqnarray}
Here the field $\Phi$ is a classical background 
field. By means of 
\begin{eqnarray}
V(x) =\lambda ' \Phi ^2  \label{22.2}
\end{eqnarray}
it defines the potential in (\ref{22.1}) for the field $\varphi (x)$, which
should be quantized in the background of $V(x)$. 
The embedding into an external system
is necessary in order to guarantee the renormalizability of the ground state
energy. 

The complete energy 
\begin{eqnarray}
E[\Phi ] =E_{class} [\Phi ] +E_{\varphi} [\Phi ]\label{E}
\end{eqnarray}
of the system consists of the classical part and the 
$1$-loop contributions resulting from the ground state energy of the quantum
field $\varphi$ in the background of the field $\Phi$. The classical part
reads
\begin{eqnarray}
E_{class}[\Phi ]=\frac 1 2 V_g +\frac 1 2 M^2 V_1 +\lambda V_2, \label{22.3} 
\end{eqnarray}
with the
definitions 
$V_g =\int d^3x (\nabla \Phi )^2$, $V_1 =\int d^3x \Phi^2$ and 
$V_2 =\int d^3x \Phi^4$. 
Here $M^2$ and $\lambda$ are the bare mass respectively coupling constant 
which need renormalization as we will explain in the following. 
For the ground state energy one defines \cite{blauvisserwipf88}
\begin{eqnarray}
E_{\varphi}[\Phi ] =\frac 1 2 \sum_{(n)} (\lambda_{(n)} ^2 +m^2 )
 ^{1/2-s} \mu^{2s}
,\label{22.4}
\end{eqnarray}
where $\mu$ is an arbitrary mass parameter and $s$ is a regularization 
parameter which has to be put to zero after renormalization. Furthermore, 
$\lambda_{(n)}$ are the eigenvalues of the corresponding wave equation
\begin{eqnarray}
(-\Delta +V(x))\phi_{(n)}(x) =\lambda _{(n)}^2 \phi_{(n)}(x) .\label{22.5}
\end{eqnarray}
For the moment we assume the space to be a large ball of radius $R$ as an 
intermediate step to have discrete eigenvalues and thus a discrete multiindex
$(n)$.

It is convenient to express the ground state energy (\ref{22.4}) in terms of
the zeta function
\begin{eqnarray}
\zeta_V (s) = \sum_{(n)} (\lambda_{(n)} ^2 +m^2 )^{-s}\label{zetascat}
\end{eqnarray}
of the wave operator with potential $V(x)$ as defined in (\ref{22.5})
by 
\begin{eqnarray}
E_{\varphi } [\Phi ] =\frac 1 2 \zeta_V (s-1/2) \mu^{2s}.\label{groundzeta}
\end{eqnarray}
In general $\zeta_V (-1/2)$ will be a divergent quantity and a renormalization
procedure for the definition of
$E_{\varphi}[\Phi ]$ is needed. It is easily discussed in terms of the 
asymptotic $t\to 0$ heat-kernel expansion associated with the wave equation
(\ref{22.5}),
\begin{eqnarray}
K(t) =\sum_{(n)} \exp\left(-\lambda_{(n)} ^2 t\right) 
&\sim& \left(\frac 1 {4\pi t}\right)^{3/2} e^{-tm^2}\sum_{j=0}^{\infty} A_j t^j. \label{heat} \\[-15pt]
&{\stackrel{t\to 0}{}}& \nonumber
\end{eqnarray}
The heat-kernel expansion in Eq.~(\ref{heat}) will also contain boundary terms
which in addition to the terms written down explicitly will lead to half integer
powers in $t$. However, ultimately we are interested in the $R\to\infty$ limit
and we will subtract the Minkowski space contribution in order to normalize 
$E_{\varphi} [\Phi =0] =0$. For the case that $V(r) \sim r^{-2-\epsilon}$ for $r\to
\infty$, $\epsilon > 0$, these terms will then disappear.

For the renormalization, only the first three terms are relevant which read
explicitly $A_0 =\int\,d^3x$,  $A_1 =-\int d^3x V(x)$
and $A_2 =(1/2) \int d^3x V^2(x)$. Their contributions to the ultraviolet
divergencies in the groundstate energy read
\begin{eqnarray}
E_{\varphi} ^{div} [\Phi ] &=& -\frac{m^4}{64\pi^2}\left(\frac 1 {s} 
+\ln\frac
{4\mu^2}{m^2} -\frac 1 2 \right) A_0 \nonumber\\ 
& &+\frac{m^2}{32\pi^2}\left(\frac 1 {s} 
+\ln\frac{4\mu^2}{m^2} -1\right) A_1
\label{endiv}\\ 
& &-\frac 1 {32\pi^2} \left(
\frac 1 {s} +\ln \frac{4\mu^2}{m^2}-2\right) A_2.\nonumber
\end{eqnarray}
The first term is a constant independent of the background field $\Phi$ and
it can be dropped (especially it will be absent after subtraction of the 
Minkowski space contribution). In a more general context of a gravitational
background field this term would yield a renormalization of the cosmological 
constant. The second term can be absorbed in a renormalization of the mass
$M$ of the background field
\begin{eqnarray}
M^2 & \to & M^2 +\frac{\lambda 'm^2}{16 \pi^2 } \left( -\frac 1 {s} +1 +\ln
\frac{m^2}{4\mu^2}\right)\label{22.7}
\end{eqnarray}
and the third one in the coupling constant $\lambda$ by
\begin{eqnarray}
\lambda & \to & \lambda +\frac{{\lambda '} ^2}{64\pi^2} \left(
-\frac 1 {s} +2 +\ln\frac{m^2}{4\mu^2}\right).\label{22.8}
\end{eqnarray}
The kinetic term $V_g$ in $E [\Phi ]$ suffers no renormalization. By
defining 
\begin{eqnarray}
E_{\varphi}^{ren} =E_{\varphi} [\Phi ] -E_{\varphi}^{div}\label{renen}
\end{eqnarray}
one obtains the finite groundstate energy, which is normalized in a way that
the functional dependence on $\Phi ^2$ present in the classical energy is now
absent in the quantum corrections $E_{\varphi}^{ren}$. 

Let us note that this is just the well known general renormalization scheme
written down here explicitly in the notations needed in our case.

\section{Scattering theory and Jost functions}
\setcounter{equation}{0}
Let us now restrict to a spherically symmetric background field $\Phi (r)$. 
Then the multiindex $(n)\to n,l,m$ consists of the main quantum number $n$, the
angular momentum number $l$ and the magnetic quantum number $m$.
In polar coordinates the ansatz for a solution of the wave equation (\ref{22.5})
reads
\begin{eqnarray}
\phi _{(n)} (x) =\frac 1 r \phi_{n,l} (r) Y_{lm}(\theta , \varphi )\label{ansatz}
\end{eqnarray}
where the radial wave equation takes the form
\begin{eqnarray}
\left[\frac{d^2 }{dr^2} -\frac{l(l+1)}{r^2} -V (r) 
+\lambda_{n,l}^2 \right]
\phi_{n,l} (r) =0. \label{2.1}
\end{eqnarray}
Now we use the standard scattering theory within $r\in [0,\infty)$ and have the momentum $p$ instead of the discrete $\lambda_{n,l}$. Let $\phi_{p,l} (r)$ 
 be  the so called 
regular solution which is defined as to have the same behaviour at $r\to 0$ as the solution without potential
\begin{eqnarray}
\phi_{p,l} (r) &\sim& j_l (pr)\label{regular}\\[-5pt]
&{\stackrel {r \to 0}{}}&\nonumber
\end{eqnarray}
with the spherical Bessel function $j_l $ \cite{taylor72}.
This regular solution defines the Jost 
function $f_l$ through its asymptotics as $r\to\infty$,
\begin{eqnarray}
\phi_{l,p}(r) &\sim& \frac i 2 \left[ f_l(p) \hat h _l^- (pr) -
        f_l^*(p) \hat h _l^+ (pr) \right],\label{2.2}\\[-5pt]
&{\stackrel{r\to\infty}{}}&\nonumber
\end{eqnarray}
where $\hat h _l^- (pr)$ 
and $\hat h _l^+ (pr)$ are the Riccati-Hankel functions \cite{taylor72}. 

Now we use the Jost function to transform the frequency sum in 
Eq.~(\ref{zetascat}) in a contour integral. Let us assume for a moment that the
support of the potential is contained in the cavity of radius $R$.
Then the above Eq.~(\ref{2.2}) gets exact at $r=R$ and may be interpreted as
an implicit equation for the eigenvalues $p=\lambda_{n,l}$.
Choosing Dirichlet boundary conditions at $r=R$, $\phi_{p,l}(R) =0$, it reads
explicitly,
\begin{eqnarray}
 f_l(p) \hat h _l^- (pR) - f_l^*(p) \hat h _l^+ (pR) =0.\label{2.3}
\end{eqnarray}
As already mentioned, 
ultimately we are interested in the limit $R\to \infty$ and in that limit
the results will not depend on the boundary condition chosen, once we assume
that $V(r) \sim r^{-2-\epsilon}$ for $r\to \infty$.

Let us now consider the ground state energy associated with the eigenvalues 
determined by (\ref{2.3}). It is convenient to represent the frequency sum 
in (\ref{22.4}) by a contour integral, the basic idea being explained in
detail in for example \cite{bek,bekg}. Using Eq.~(\ref{2.3}), one immediately
finds 
\begin{eqnarray}
E_\varphi [\Phi ] &=& \mu^{2s}\sum_{l=0}^{\infty}(l+1/2) \int\limits_{\gamma}\frac{dp}{2\pi i}\,\,(p^2+m^2)^{1/2-s}\frac{\partial}{\partial p} 
               \ln [ f_l(p) \hat h _l^- (pR) - f_l^*(p) \hat h _l^+ (pR)] \nonumber\\ 
& &+\mu^{2s}\sum_{l=0}^{\infty}(l+1/2) \sum_n (m^2-\kappa_{n,l}^2)^{1/2-s},\label{2.4}
\end{eqnarray}
with $-\kappa_{n,l}^2$ as the energy eigenvalues of the bound states
with given orbital momentum $l$. The contour
$\gamma$ is chosen counterclockwise enclosing 
all real solutions of Eq.~(\ref{2.3})
on the positive real axis. 
The division of the discrete eigenvalues within the large ball into 
positive (inside $\gamma$) and negative ones ($-\kappa_{n,l}^2$), is determined 
by the conditions that $V(r) \to 0$ for $r\to\infty$. In that way, in the 
limit of the infinite space the negative eigenvalues become the usual
boundstates and the $\lambda_{n,l} > 0$ turn into the scattering states.

For the calculation of (\ref{2.4}), 
as the next step one deforms the contour $\gamma$ 
to the imaginary axis. A contour coming from $i \infty +\epsilon $, crossing the imaginary
axis at some positive value smaller than the smallest $\kappa _n$ and going to
$i\infty -\epsilon$ results first. Shifting the contour over the bound state values
$\kappa_n $ which are the zeroes of the Jost function on the
imaginary axis, the bound state 
contributions in Eq.~(\ref{2.4}) are cancelled and in the limit $R\to 
\infty$, subtracting the Minkowki space contributions, one finds
\begin{eqnarray}
E_\varphi [\Phi ]  =-{\cos \pi s\over \pi}\mu^{2s} \sum_{l=0}^{\infty}(l+1/2)  
\int\limits_{m}^{\infty}dk\,\, [k^2-m^2]^{\frac{1}{2}-s}~\frac{\partial}{\partial k}\ln  f_l (ik).\label{2.5}
\end{eqnarray}
This is the representation of the ground state energy (and by means of 
(\ref{groundzeta}) as well of the zeta function)
 in terms of the Jost function,
which is the starting point of our following analysis. It has the nice 
property, that the dependence on the bound states is not present explicitly,
being however contained in the Jost function by its properties on the positive
imaginary axis. To the authors knowledge, this 
representation of the ground state energy respectively for the zeta 
function is not known in the 
literature. Its analog proved to be useful for explicit numerical calculations 
already in the case of a potential depending only on one coordinate 
\cite{bordag95}. It is connected with the more conventional representations
by means of the analytic properties of the Jost function. Expressing
them by the dispersion relation
\begin{eqnarray}
f_l (ik) =\prod_n\left(1-\frac{\kappa_{n,l}^2}{k^2}\right) \exp\left(
-\frac 2 {\pi}\int_0^{\infty}\frac{dq\,\, q}{q^2 +k^2} \delta_l (q)\right),
\nonumber
\end{eqnarray}
where $\delta_l (q)$ is the scattering phase, we obtain from (\ref{2.5}) 
\begin{eqnarray} E_{\varphi} [\Phi ] =\mu^{2s}\sum_{l=0}^\infty \left(l+\frac{1}{2}\right)
\left\{- \sum_n \left( m^{1-2s}-\sqrt{m^2-\kappa_{n,l}^2}^{\,1-2s}\right)\right.\label{Eun}\\ 
\left.-\frac{1-2s}{\pi}\int_0^{\infty}dq\,\, {q\over\sqrt{q^2+m^2}^{1-2s}}\,\delta_l (q)\right\},\nonumber\end{eqnarray}
which gives the expression of the ground state energy through the  scattering
phase. From here one can pass to the representation through the mode density by
 integrating by part.
Note, that this representation can be obtained also directly from 
(\ref{2.4}) in the limit $R\to \infty$ by deforming the contour
$\gamma$.

Let us add a discussion on the sign of the ground state energy. In (\ref{Eun})
the first contribution results from the bound states and is completely
negative. The second contribution which containes the scattering phase
$\delta_l(q)$ is positive (negative) for an attractive (repulsive) potential,
i.e., for $V(r)<0$ ($V(r)>0$) for all $r$ \footnote{This is a well known fact
from potential scattering, see e.g. \cite{taylor72}.}. So the regularized
(still not renormalized) ground state energy $E_{\varphi} [\Phi ]$ (\ref{Eun})
is positive for a potential which is repulsive for all $r$ (there are no bound states in this case) and it is negative for a potential which is attractive for
all $r$. 
Now, if we perform the renormalization in accordance with (\ref{22.7}) and
(\ref{22.8}),
we obtain
\begin{eqnarray}
E^{ren}=E_{\varphi} [\Phi ]+{m^2\over
8\pi}\left(\frac{1}{s}+\log\frac{4\mu^2}{m^2}-1\right)\int_0^\infty{\rm
d}r\,r^2\,V(r)\nonumber \\ 
+{1\over 16\pi}\left(\frac{1}{s}+\log\frac{4\mu^2}{m^2}-2\right)\int_0^\infty{\rm
d}r\,r^2\,V(r)^2\,.\label{Eren}\end{eqnarray}
The contribution containing $A_0$ had been already subtracted in $E_{\varphi}
[\Phi ]$ (\ref{2.5}). This expression is finite for $s\to 0$, i.e., when
removing the regularization. But due to the subtracted terms there is no
longer any definite result on the sign. Note, that this is in contrast to the
case of a one-dimensional potential where it had been possible to express the
subtracted terms through the scattering phase \cite{bordag95}.

\section{Uniform asymptotics of the Jost function}
Let us continue with a detailed analysis of the ground state energy, 
Eq.~(\ref{2.5}).
As we have easily seen using heat-kernel techniques 
the non-renormalized vacuum energy $E_{\varphi} [\Phi ]$
contains divergencies in $s=0$, see Eq.~(\ref{endiv}), which are removed
by the renormalization prescription given in Eqs.~(\ref{22.7}) --
(\ref{renen}), resp. (\ref{Eren}).
The poles present are by no means obvious in the representation (\ref{2.5}) 
of $E_\varphi [\Phi]$. 
However, in order to actually perform the renormalization,
Eq.~(\ref{renen}), it is necessary to represent the groundstate energy
Eq.~(\ref{2.5}) in a form which makes the explicit subtraction of the 
divergencies visible. This will be our first task.

As is known from general zeta function theory as well as one sees 
from simply counting 
the large momentum behaviour, the representation 
Eq.~(\ref{2.5}) of $E_\varphi [\Phi ]$ will be convergent for $\,\mbox{Re}\, s >  2$. 
However, for the calculation of the ground
state energy we need the value of Eq.~(\ref{2.5}) in $s=0$, thus an 
analytical continuation to the left has to be constructed.
The basic idea is the same as the one presented in \cite{bek,bekg}: 
adding and subtracting the leading uniform asymptotics of the 
integrand in Eq.~(\ref{2.5}). Let
\begin{eqnarray}
E_\varphi [\Phi ]=E_f+E_{as} \label{3.3a}
\end{eqnarray}
where 
\begin{eqnarray}
 E_f =-\frac{\cos (\pi s)}{\pi} \mu^{2s} 
\sum_{l=0}^{\infty}(l+1/2) \int\limits_{m}^{\infty}dk\,\, [k^2-m^2]^{\frac{1}{2}-s}\frac{\partial}{\partial k} [\ln f_l (ik) - \ln f_l ^{asym} (ik) ]   
\label{ns}
\end{eqnarray}
and 
\begin{eqnarray}
 E_{as} =-\frac{\cos (\pi s)}{\pi} \mu^{2s} \sum_{l=0}^{\infty}(l+1/2) \int\limits_{m}^{\infty}dk\,\, [k^2-m^2]^{\frac{1}{2}-s}\frac{\partial}{\partial k} \ln f_l ^{asym} (ik).\label{as}
\end{eqnarray}
The idea is, that as many asymptotic terms are subtracted as to 
allow to put $s=0$ in 
the integrand of $E_f$. This term will then (in general) be evaluated
numerically.  In $E_{as}$ the analytic continuation to $s=0$ can be 
done explicitly showing that the pole contributions cancel
when subtracting $E_\varphi^{div}[\Phi ]$, Eq.~(\ref{endiv}). 
Note, that the contribution 
resulting from $A_0$ has been dropped already in Eq.~(\ref{2.5}).

The first task thus is to obtain the asymptotics of the Jost functions.
This may be done by using the integral equation 
(Lippmann-Schwinger equation) known from scattering
theory \cite{taylor72}. For the Jost function one has ($\nu \equiv l+1/2$) 
\begin{eqnarray} 
f_l (ik) =1+\int\limits_0^{\infty} dr\,\,r\,V(r) \phi_{l,ik} (r)
   K_{\nu} (kr),\label{3.1}
\end{eqnarray}
with the regular solution given by the integral equation
\begin{eqnarray}
\phi_{l,ik} (r) &=& I_{\nu} (kr) +
\int\limits_0^r dr'\,\, r'\,\,[ I_{\nu } (kr) K_{\nu} (kr')-
 I_{\nu} (kr') K_{\nu} (kr)] V(r') \phi_{l,ik}(r'). \label{3.2}
\end{eqnarray}
General zeta function theory tells us that the divergence at $s=0$ contains
at most terms of order $V^2$. Thus one might expand $\ln f_l (ik)$ in powers
of $V$ and take into account only the asymptotics of terms up to $ {\cal O} (V^2)$.
The expansion in powers of $V$ is
easily obtained. Using Eqs.~(\ref{3.1}) and (\ref{3.2}) one finds
\begin{eqnarray}
\ln f_l (ik) &=& \int\limits_0^{\infty}dr\,\, r V(r) K_{\nu} (kr) I_{\nu} (kr)\nonumber\\ 
& &-\int\limits_0^{\infty}dr\,\, r V(r) K_{\nu}^2 (kr) \int\limits_0^rdr'\,\, r' V(r') I_{\nu}^2 (kr')\nonumber\\ 
& &+ {\cal O} (V^3). \label{3.3}
\end{eqnarray}
Now the uniform asymptotics for $l\to \infty$ of $\ln f_l (ik)$ is
essentially 
reduced to the well 
known uniform asymptotics of the modified Bessel functions $K_{\nu}$
and $I_{\nu}$, (9.7.7) and (9.7.8) in \cite{abramowitzstegun72}.
With the notation  $t=1/\sqrt{1+(kr/\nu)^2}$ and
$\eta (k) =\sqrt{1+(kr/\nu)^2}+\ln [(kr/\nu) /(1+\sqrt{1+(kr/\nu)^2})]$,
one finds for $\nu\to\infty$, $k\to\infty$ with $k/\nu $ fixed,
\begin{eqnarray}
I_{\nu} (kr   ) K_{\nu} (kr ) &\sim &
\frac 1 {2\nu t} +\frac{t^3}{16\nu^3}\left( 1-6 t^2 
+5 t^4 \right) + {\cal O} (1/\nu^4 )\nonumber\\ 
I_{\nu} (k r' ) K_{\nu} (kr)& \sim&
\frac 1 {2\nu} \frac{e^{-\nu (\eta (k) -\eta (kr'/r))}}
{(1+(kr/\nu)^2)^{1/4} (1+(kr'/\nu)^2)^{1/4}}
\left[ 1+ {\cal O} (1/\nu )\right].\nonumber
\end{eqnarray}
Using these terms in the rhs of Eq.~(\ref{3.3}) we define
\begin{eqnarray} 
\ln f_l^{asym}(ik)&=&
 \frac 1 {2\nu} \int\limits_0^{\infty}dr\,\,
\frac{r\,V(r)}{\left[1+\left(\frac {kr} {\nu}\right)^2 \right]^{1/2}} \nonumber\\ 
& &+\frac 1 {16\nu^3}\int\limits_0^{\infty}dr\,\,
\frac{r\,V(r)}{ \left[1+\left(\frac {kr} {\nu}\right)^2 \right] ^{3/2}}\left[1 - \frac 6  
  {\left[1+\left(\frac {kr} {\nu}\right)^2 \right] }
+ \frac 5  {\left[1+\left(\frac {kr} {\nu}\right)^2 \right]^{2}}\right]\nonumber\\ 
& & -\frac 1 {8\nu^3}\int\limits_0^{\infty}dr\,\,\frac{r^3\, V^2 (r)}
{ \left[1+\left(\frac {kr} {\nu}\right)^2 \right]^{3/2}}. 
\label{3.3b}
\end{eqnarray}
Thereby the $r'$-integration 
in the term quadratic in $V$ has been performed
by the saddlepoint method using the monotony of $\eta (k)$.
Now by means of (\ref{3.3b})  the limit $s\to 0$ can be performed in 
Eq. (\ref{ns}) and we obtain
\begin{eqnarray}
 E_f =-\frac{1}{\pi} \sum_{l=0}^{\infty}(l+1/2) {\int\limits_{m}^{\infty}dk\,\, \sqrt{k^2-m^2}}
\frac{\partial}{\partial k}\left(\ln f_l (ik) -\ln f_l^{asym} (ik)
\right) ,
\label {efin}
\end{eqnarray}
a form which is suited for a numerical evaluation.

For 
$E_{as}$ at $s=0$ one might explicitly find
the analytical 
continuation. First of all, the $k$-integrals may be done 
using
\begin{eqnarray}
{\int\limits_{m}^{\infty}dk\,\, [k^2-m^2]^{\frac{1}{2}-s}\frac{\partial}{\partial k}}\left[1+\left(\frac {kr} {\nu}\right)^2 \right] ^{-\frac{n}{2}}=
 -\frac{\Gamma (s+\frac{n-1}{2})
\Gamma (\frac{3}{2}-s)}{ \Gamma (n/2)}
 \frac{\left(\frac{\nu}{mr}\right)^n
m^{1-2s}}{\left(1+
\left(\frac{\nu}{mr}\right)^2\right)^{s+\frac{n-1}{2}}}\label{kint}
\end{eqnarray}
to 
yield Eq.~(\ref{a1}) in the appendix A.
>From this, the renormalization, Eq.~(\ref{renen}) can be carried out
and we arrive at
\begin{eqnarray}
E_{as}-E_{\varphi}^{div}=
\frac 1 {48\pi} \int\limits_0^{\infty} dr\,\, V(r) 
\left[6m^2 r^2 -12 \zeta_R' (-1) -3\ln 2 \right.\hfill\nonumber\\ 
\qquad \left.-6r^2 \ln (16 m^2 r^2) \left(m^2 +\frac 1 2 V(r) \right) 
-\gamma (1+12 m^2 r^2 +6 r^2 V(r) )\right]+E_{as}^{sum}.\nonumber
\end{eqnarray}
The complete energy $E[\Phi]$, (\ref{E}), 
consisting of $E_{class}[\Phi]$, (\ref{22.3}), of the background 
field $\Phi$  and of the renormalized loop contribution 
(\ref{renen}) 
\begin{eqnarray} E_\varphi ^{ren}=E_{\varphi}[\Phi]-
E_{\varphi}^{div}=E_f+E_{as}-E_{\varphi}^{div}\label{4.9a}\end{eqnarray}
reads
\begin{eqnarray}
E[\Phi ]&=& \frac 1 2 V_g +\frac 1 2 M_{ren} ^2 V_1 +\lambda _{ren} V_2
\label{energie}\\ 
& &+\frac 1 {48\pi} \int\limits_0^{\infty} dr\,\, V(r) 
\left[6m^2 r^2 -12 \zeta_R' (-1) -3\ln 2 \right.\nonumber\\ 
& & \qquad \left.-6r^2 \ln (16 m^2 r^2) \left(m^2 +\frac 1 2 V(r) \right) 
-\gamma (1+12 m^2 r^2 +6 r^2 V(r) )\right]\nonumber\\ 
& & +E_f+E_{as}^{sum}. \nonumber
\end{eqnarray}
The remaining task for the analysis of the ground state energy in the presence
of a spherically symmetric potential is the numerical analysis of the above 
quantity. To achieve that, a (as a rule numerical) knowledge of the Jost 
function $f_l (ik)$ is necessary. Apart from this only integrals of the 
potentials and convergent sums have to be dealt with which presents no problem.

\section{Example: Square-well potential}
If the potential has a compact support, there is a formalism how to
obtain the Jost function (at least in principle). Starting point is
the observation, that one may write the regular solution in the form
\begin{eqnarray}
\phi_{l,p}(r) =u_{l,p} (r) \Theta (R-r) +\frac i 2 
\left[ f_l(p) \hat h _l^- (pr) - f_l^*(p)
 \hat h _l^+ (pr) \right] \Theta (r-R),\label{4.1}
\end{eqnarray}
where $V(r)=0$ for $r\geq R$ is assumed. The matching conditions then read
\begin{eqnarray}
 u_{l,p} (R) &=& \frac i 2 \left[ f_l(p) \hat h _l^- (pR) - f_l^*(p) \hat h _l^+ (pR)\right],\nonumber\\ 
 u_{l,p}' (R) &=& \frac i 2 p \left[ f_l(p)  {\hat h}_l^{- \prime} (pR) -  f_l^*(p)  {\hat h} _l^{+ \prime} (pR) \right].\nonumber
\end{eqnarray}
Combining the two equations and using the fact that the Wronskian 
determinant of $\hat h _l^{\pm}$ is $2i$, one arrives at
\begin{eqnarray}
 f_l(p) = -\frac 1 p \left( p u_{l,p} (R)  {\hat h} _l^{+ \prime} (pR) - u_{l,p}' (R) \hat h _l^+ (pR) \right),\label{4.2}
\end{eqnarray}
which gives the expression of the Jost function for a potential with a
compact support through the wave function.

For the square-well potential $V(r) =V_0 \Theta \,(R-r)$ it is easily seen that
\begin{eqnarray}
u_{l,p} (r) = \left(\frac p {\tilde q } \right) ^{l+1} \hat j _l (\tilde q r)
\nonumber
\end{eqnarray}
with $\hat j _l $ the Riccati-Bessel function and $\tilde q =\sqrt{p^2-V_0}$.
 
So we obtain the well known formula
\begin{eqnarray}
f_l ^{sw} (ik) = R \left( \frac k q \right) ^{\nu} [q I_{\nu}
' (qR) K_{\nu} (kR) -kI_{\nu} (qR) K_{\nu} ' (kR) ],\label{4.3}
\end{eqnarray}
with $q=\sqrt{k^2+V_0}$.
This has to be used for the numerical evaluation of 
Eq.~(\ref{energie}).
Due to the simple form of the potential $V(r)$ the $r$-integrals may be
done explicitly and numerically easy tractable expressions result. 
For $E_f$ we obtain
\begin{eqnarray}
E_f &=&  -\frac 1 {\pi} \sum_{l=0}^{\infty}(l+1/2)  \int\limits_{m}^{\infty}dk\,\, [k^2-m^2]^{1/2} \frac{\partial}{\partial k}
        \times\label{4.5}\\ 
& &\left\{ \ln f_l ^{sw} (ik) -\frac{V_0 \nu} {2k^2} \left(\left[1+\left(\frac {kR} {\nu}\right)^2 \right] ^{1/2}-1
\right)
\right.\nonumber\\ 
& &+\frac{V_0}{16\nu k^2}\left[ \left[1+\left(\frac {kR} {\nu}\right)^2 \right] ^{-1/2} - 2\left[1+\left(\frac {kR} {\nu}\right)^2 \right] ^{-3/2}+
 \left[1+\left(\frac {kR} {\nu}\right)^2 \right] ^{-5/2}\right]  \nonumber\\ 
& &\left.+\frac{V_0^2}
{8\nu k^2 }\left[\frac{R^2+2(\nu /k)^2}{\left(
1+(kR/\nu)^2 \right)^{1/2}} -
2\left(\frac{\nu} k \right)^2\right]\right\},\nonumber
\end{eqnarray}
which is, as one might easily check, a finite expression.
For the renormalized contributions of the asymptotic terms, we obtain
\begin{eqnarray}
E_{as}-E_{\varphi}^{div}&=&\frac{V_0R}{48\pi}\left[-12\zeta_R'(-1)+
\frac{10} 3 m^2R^2 -3\ln 2 -2m^2R^2\ln (16m^2R^2)\right.\label{squasym}\\ 
& &\left. +\frac 2 3 V_0R^2 -V_0 R^2 \ln (16m^2R^2) -\gamma (1+4m^2 R^2 
+2V_0 R^2)\right]+E_{as,sw}^{sum},\nonumber
\end{eqnarray}
where $E_{as,sw}^{sum}$ is given in appendix B. It is presented in form of
quickly convergent series which allow an easy numerical analysis.

\unitlength1cm
\begin{figure}[h]
\begin{picture}(15,12)
\epsfxsize=15cm
\put(0,-4){\epsffile{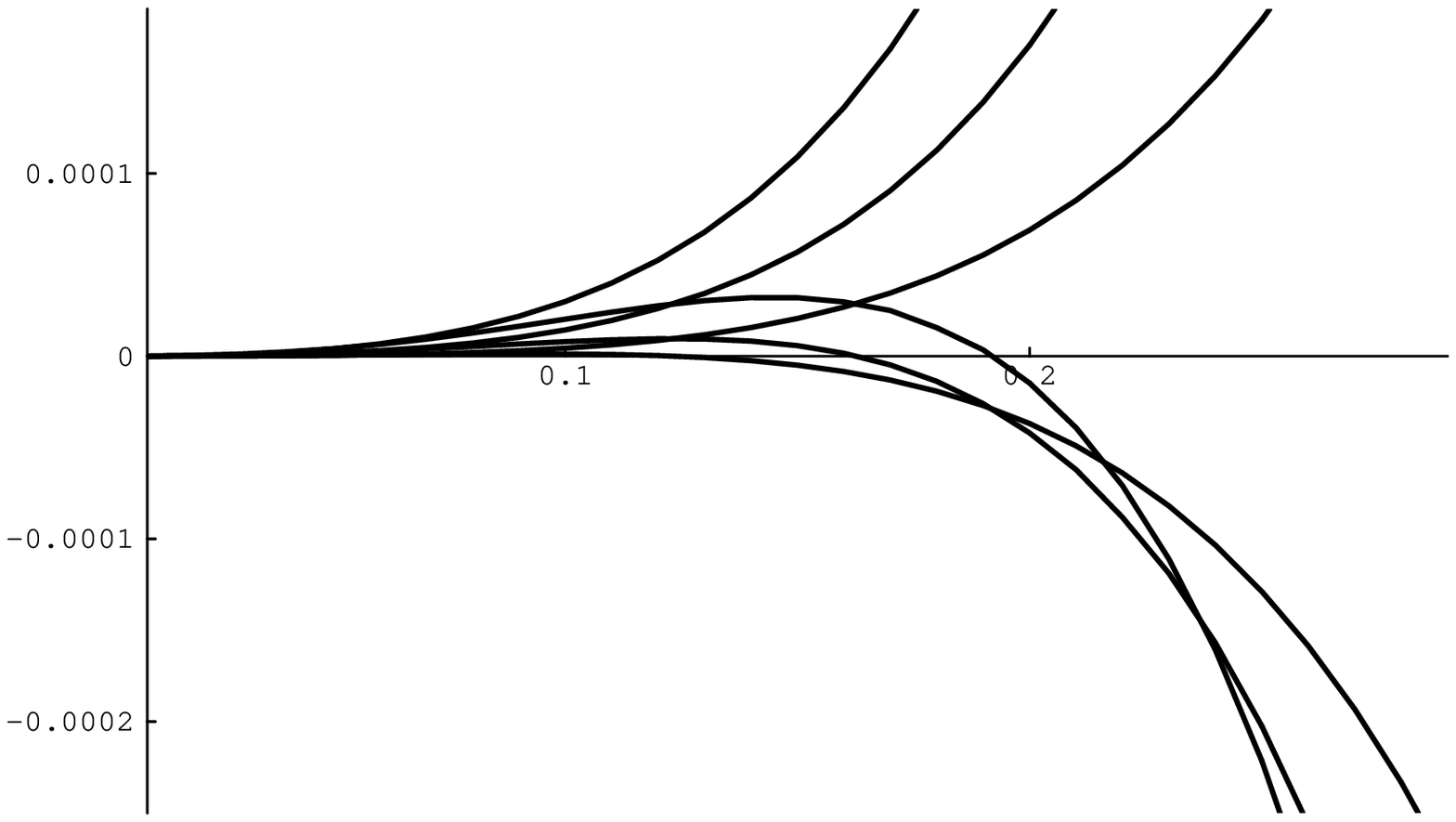}}
\put(13,10.7){\hbox{\footnotesize $V_0=0.1$}}
\put(10.8,10.7){\hbox{\footnotesize $V_0=0.4$}}
\put(9,10.7){\hbox{\footnotesize $V_0=0.9$}}
\put(14.02,3.8){\hbox{\footnotesize $V_0=-0.1$}}
\put(13.5,1.8){\hbox{\footnotesize $V_0=-0.4$}}
\put(11.2,2.83){\hbox{\footnotesize $V_0=-0.9$}}
\put(15.3,6.9){\hbox{\footnotesize $R$}}
\put(1.0,10.8){\hbox{\footnotesize $E_{\varphi}^{ren} [\Phi ]$}}
\put(0,1){Fig. 1~~~\parbox[t]{14cm}{The complete energy as a function of the
radius $R$ for different values of the height of the potential wall $V_0$ for $m=1$.}}
\end{picture}
\end{figure}

The results for $E_{\varphi}^{ren} [\Phi ]$ are presented in Fig.~1.
For small values of the radius $R$ of the support of the potential, it is 
seen that even for negative values of the potential, $V_0<0$, a 
positive vacuum energy results. The reason is, that for small values of $R$
the bound state energies are located closely to zero, giving only small
negative contributions. For increasing $R$, their number and their values
are increasing leading necessarily to negative vacuum energies.

The behaviour described is the one we expected from the one-dimensional results
presented in \cite{bordag95}. However, the absolute orders of magnitude are by
two lower than in the corresponding one-dimensional considerations.

\section{Conclusions}
In this article we reduced the task of calculating the vacuum energy of a 
scalar field in the presence of a spherically symmetric background field to
the corresponding quantum mechanical scattering problem. We were able to
present the renormalized vacuum energy solely in terms of the 
quantum mechanical scattering dates summarized by the Jost function. For the 
example of the square-well potential we showed that a direct numerical analysis 
of the vacuum energy is possible.

Several extensions of our approach are necessary. From the physical point of 
view the consideration of higher-spin fields is necessary and envisaged.
In addition, in order to apply our formalism to classical background fields
like f.e.~sphalerons and electroweak skyrmions, a numerical analysis of 
Eq.~(\ref{efin}) is necessary for cases when the Jost function is known
only numerically. However, the scattering theory developed during the last
decades provides many techniques and results so that also here progress 
seems possible.

\section*{Acknowledgments}
We thank Emilio Elizalde and Sergio Leseduarte for interesting and helpful
discussions. Furthermore, Sergio Leseduarte is warmly thanked for his
advice in the numerical analysis for the square well potential presented
in the article. Finally discussions with Kimball Milton during the 
conference on {\it Quantum field theory under the influence of external conditions} (Sept., 18-22, 1995) in Leipzig
on how to obtain analytical bounds on the numerical 
errors in our analysis  are gratefully acknowledged.

This investigation has been supported  by the DFG under the contract number 
Bo 1112/4-1.

\begin{appendix}
\renewcommand{\theequation}{{\mbox A}.\arabic{equation}}
\setcounter{equation}{0}
\section{Contributions of the asymptotic terms to the zeta function}

Here we calculate the analytic continuation of $E_{as}$ (\ref{as}) using the asymptotics of the Jost functions $\ln f_l^{as}(ik)$ (\ref{3.3b}).
After carrying out the $k$-integration by means of (\ref{kint}) we obtain
\begin{eqnarray}
E_{as}& =& -\frac{\Gamma (s)\mu^{2s}}{2\sqrt{\pi}m^{2s}
        \Gamma (s-1/2)} \int\limits_0^{\infty}dr\,\, V(r) \sum_{l=0}^{\infty} 
                \nu \left[1+\left(\frac{\nu}{mr}\right)^2 \right] ^{-s}\nonumber\\ 
&& -\frac{m^{1-2s}\mu^{2s}}{2\sqrt{\pi} \Gamma (s-1/2)} \int\limits_0^{\infty}dr\,\, r V(r)\sum_{l=0}^{\infty} 
\left\{\frac{\Gamma (s+1)}{4(mr)^3} \nu\left[1+\left(\frac{\nu}{mr}\right)^2 \right] ^{-s-1}\right.\nonumber\\ 
& &-\frac{\Gamma (s+2)} {(mr)^5} \nu^3 \left[1+\left(\frac{\nu}{mr}\right)^2 \right] ^{-s-2}
\left. +\frac{\Gamma (s+3)}{3(mr)^7}\nu^5 \left[1+\left(\frac{\nu}{mr}\right)^2 \right] ^{-s-3}\right\},\nonumber\\ 
&& +\frac{\Gamma (s+1)\mu^{2s}}{4\sqrt{\pi}m^{2s+2}\Gamma (s-1/2)}
\int\limits_0^{\infty} dr\,\, V^2 (r)\sum_{l=0}^{\infty} \nu \left[1+\left(\frac{\nu}{mr}\right)^2 \right] ^{-s-1}.\label{a1}
\end{eqnarray}
The analytical continuations of the above expressions to the value
$s=0$ is easily obtained and listed below,
\begin{eqnarray}
E_{as} &=& \frac 1 {96\pi}  \int\limits_0^{\infty}dr\,\,V(r) (1-12m^2r^2)\left(\frac 1 {s}+\ln r^2\mu^2\right) \nonumber\\ 
& &+\frac 1 {48\pi}\int\limits_0^{\infty}dr\,\,V(r) 
\left[-1-12\zeta_R'(-1)+12m^2r^2(1-\gamma-3\ln 2)\right]\nonumber\\ 
&&-\frac 1 {96\pi}  \int\limits_0^{\infty}dr\,\,V(r) \left(\frac 1 {s}+\ln r^2\mu^2\right)\label{a5}
+\frac 1 {48\pi} \int\limits_0^{\infty}dr\,\,V(r) [1-\gamma-3\ln 2]\nonumber\\ 
&&-\frac 1 {16\pi} \int\limits_0^{\infty}dr\,\,r^2 V^2(r)\left(\frac 1 {s}+\ln r^2\mu^2\right)
+\frac 1 {8\pi} \int\limits_0^{\infty}dr\,\,r^2 V^2(r) (1-\gamma -3\ln 2)\nonumber\\ 
& &+E_{as}^{sum}+ {\cal O} (s),\label{a2}\\ 
\nonumber\end{eqnarray}
where the sums in
\begin{eqnarray}
E_{as}^{sum}&=&\frac 1 {4\pi}\int\limits_0^{\infty}dr\,\,V(r) \sum_{l=0}^{\infty} \nu \left[\left(\frac
{mr}{\nu}\right)^2  -\ln\left(1+ \left(\frac {mr}{\nu}\right)^2\right)
\right]\nonumber\\ 
&&+\frac 1 {4\pi} \int\limits_0^{\infty}dr\,\,V(r) 
\sum_{l=0}^{\infty} \frac 1 {\nu} \left\{\frac 1 4 \left[1+\left(\frac{mr}{\nu}\right)^2 \right]^{-1} -
 \left[1+\left(\frac{mr}{\nu}\right)^2 \right]^{-2}\right.\nonumber\\ 
& &\left.   +\frac 2 3 \left[1+\left(\frac{mr}{\nu}\right)^2 \right]^{-3} +\frac 1
 {12}\right\}\nonumber\\ 
&&-\frac 1 {8\pi} \int\limits_0^{\infty}dr\,\,r^2 V^2(r) \sum_{l=0}^{\infty} \frac 1 {\nu} \left( \left[1+\left(\frac{mr}{\nu}\right)^2 \right]
^{-1} -1\right).\label{Esum}
\end{eqnarray}
converge.

This expressions have been the basis to give the results for the special
example listed in appendix B.   
      
\section{Asymptotic contributions for the square well potential}
In this appendix we give the result for $E_{as,sw}^{sum}$ which reads
\begin{eqnarray}
E_{as,sw}^{sum}
&=&
\frac{V_0R}{4\pi} \sum_{l=0}^{\infty}\nu\left[\frac 1 3 \left(\frac{mR}{\nu}\right)^2 -
\ln\left(1+\left(\frac{mR}{\nu}\right)^2 \right) +2-\frac{2\nu}{Rm} \arctan\left(\frac
{Rm}{\nu}\right)\right]\nonumber\\ 
& &+\frac{V_0R}{4\pi} \sum_{l=0}^{\infty} \frac 1  {\nu} \left[ 
-\frac 1 4 \left(1+\left(\frac{mR}{\nu}\right)^2\right)^{-1} +\frac 1 6 \left(1+\left(\frac{mR}{\nu}\right)^2\right)^{-2}
+\frac 1 {12}\right]
\nonumber\\ 
& &-\frac{V_0^2 R^3}{8\pi}\sum_{l=0}^{\infty} \frac 1 {\nu} \left\{
 \left(\frac{\nu}{mR}\right)^2 
 -\left(\frac{\nu}{mR}\right)^3\arctan\left(\frac{mR}
{\nu}\right) -\frac 1 3 \right\}.\nonumber
\end{eqnarray}
These sums have been used for the numerical analysis of the groundstate 
energy in the presence of a square-well potential 
\end{appendix}

\end{document}